\documentclass[sigconf]{acmart} 
\include{graphicsx}
\AtBeginDocument{%
  \providecommand\BibTeX{{%
    \normalfont B\kern-0.5em{\scshape i\kern-0.25em b}\kern-0.8em\TeX}}}

\setcopyright{acmcopyright}

\acmConference[UNPUBLISHED]{}{}{}
\acmBooktitle{UNPUBLISHED} 

\begin{document}

\title{Insights for post-pandemic pedagogy across one CS department}

\author{Maxwell Bigman, Yosefa Gilon, Jenny Han, John Mitchell}
\email{{mbigman, ygilon, jcm}@stanford.edu, { jennyhan}@cs.stanford.edu}


\renewcommand{\shortauthors}{}
\newcommand\todo[1]{\textcolor{red}{#1}}


\begin{abstract}
Adaptive remote instruction has led to important lessons for the future, including rediscovery of known pedagogical principles in new contexts
and new insights for supporting remote learning.
Studying one computer science department that serves residential and remote undergraduate and graduate students, we conducted interviews with stakeholders in the department (n=26) and ran a department-wide student survey (n=102) during the four academic quarters from spring 2020 to spring 2021. Our case study outlines what the instructors did, summarizes what instructors and students say about courses during this period, and provides recommendations for CS departments with similar scope going forward. Specific insights address: (1) how instructional components are best structured for students; (2) how students are assessed for their learning; and (3) how students are supported in student-initiated components of learning.  The institution is a large U.S. research university that has a history of online programs including online enrollment in regular on-campus courses and
large-scale open enrollment courses. 
Our recommendations to instructors across the scope of this department
may also be applicable to other institutions that provide technology-supported in-person instruction, remote enrollment, and hybrid courses combining both modalities.
\end{abstract}

\ccsdesc[300]{Applied computing~Distance learning}
\ccsdesc[500]{Applied computing~Interactive learning environments}

\keywords{CS education, remote learning, CS pedagogy
CS departments, COVID-19, pandemic}


\maketitle

\section{Introduction}
Colleges and computer science departments worldwide creatively responded to the pressing need for remote instruction spurred by the global pandemic and its consequences. As students and student expectations have been altered by a traumatic and life-changing period, instructors have been open-minded and resourceful in serving students successfully while confronting their own challenges.

We believe several cultural and pedagogical changes spurred by the pandemic will provide better learning opportunities for students as they return in hybrid or in full. 
Hoping to help important shifts take hold on a department-wide rather than course-by-course basis, we conducted a longitudinal best-effort inventory of pandemic-era CS course experiences, combined with ongoing department-wide pedagogical support and information exchange.
While significant changes have been observed at many institutions \cite{siegel2021teaching}, 
the CS department represented in this case study both serves approximately 20\% of residential students and has a strong background in online courses.
Housed in a large research university with a computer science teaching program at both undergraduate and graduate levels, a significant remote student population enrolls in the same on-campus classes as residential students. Over two decades, hybrid enrollment in many larger classes led to widespread residential use of asynchronously accessible video. 
Many instructors were also previously active in the MOOC movement from 2011 onward.
Thus both instructors and students were familiar with many tools of remote instruction, including pre-recorded video, video recording of synchronous class sessions, and 
online platforms that support submission, staff grading, and automated grading of student work.
Nonetheless, the pandemic led to many shocking realizations.

Perhaps surprisingly, the instructors with online teaching experience who approached pandemic instruction with the view that “we know how to do this” encountered some of the greatest difficulties. Those who approached remote instruction of displaced students as a new design problem appear to have fared best, especially those who took the most care to monitor student conditions and adapt their teaching accordingly. Looking forward from a disruptive two-year period, our primary question is:  
\textbf{What have we learned that can improve our future programs, for enrolled students on and off campus?}

In overview, Section 2 describes related work and Section 3 details our approach. In Section 4, we offer a retrospective of teaching innovations from 2020 and 2021, organized into three main categories: (1) new formats for \textbf{instructor-directed learning}, (2) alternative types of \textbf{student assessment}, and (3) \textbf {student-initiated learning}, such as office hours or peer learning. 
This section also highlights key changes in student expectations. In Sections 5 and 6, we shift toward the future, suggesting trajectories for post-pandemic teaching and recommendations for each of the three categories of activities highlighted in the paper.

\section{Related Work}
Scholarship around online teaching and learning took off in the previous decade with the advent of MOOCs, followed by a period of initial disappointment as studies suggested online learning was not scaling democratically nor reaching levels of engagement comparable to face-to-face instruction \cite{kizilcec2013deconstructing}. When the COVID-19 pandemic provided a new impetus to revisit online CS pedagogy \cite{kizilcec2021pandemic}, instructors explored interactive, human-centered online teaching practices \cite{latulipe2021cs1,piech2021code}, rethought the format and integrity of high-stakes assessments \cite{ko2021revolutionizing}, and redesigned their courses with student input \cite{bai2020lessons}.
A recent ITiCSE working group report underscored the fact that we will likely never return to a pre-pandemic educational landscape \cite{siegel2021teaching}. 
Because that working group provides an extensive
literature review, we refer the reader to their report \cite{siegel2021teaching}
rather than provide a less extensive review in the limited space 
available here.


\section{Approach}
Our case study for the purpose of instructional improvement focused on four remote-learning academic quarters, Spring 2020 through Spring 2021 (inclusive). We carried out semi-structured interviews with stakeholders in the department, tabulated course material, and conducted a department-wide student survey in June 2021. 

In total, we interviewed 19 instructors, 4 student teaching assistants (TAs) or peer advisers, and representatives from 3 undergraduate interest groups (Women in CS, American Indian Science and Engineering Society, Society of Latinx Engineers). We sent out a general call to the department and selected interviews for breadth among the instructors and TAs who volunteered their time. 
We also conducted follow-up longitudinal interviews with 6 of the instructors who had the most varied sets of insights and innovations.

The courses surveyed typically see enrollment of 50-400 students in any term, including a small number of students (5--10\%) who would have been remote regardless of the pandemic. Notably, the courses represented various tracks within the department, ranged from introductory core classes such as CS1 to large-enrollment artificial intelligence classes to smaller human-computer interaction design studios. In addition, we sought out interviews with student groups, student TAs, and peer advisers, who advocated on behalf of a diverse set of students in CS classes. 

Semi-structured interviews lasted 40--60 minutes, depending on interviewee interest. Questions were asked regarding course logistics, pedagogical experiments, ongoing challenges, and initial reactions from students. The final section of each interview followed the interest of the interviewee.

The department-wide survey to solicit students’ experiences with remote learning received 102 participants. All undergraduate and graduate students who had taken at least once CS class between Spring 2020 and Spring 2021 were invited to fill out the survey, which was disseminated via email listservs. The survey consisted of 5 Likert-scale or multiple choice questions, with optional space to elaborate, and 2 two open response questions. We made clear that participation was voluntary, optional, confidential, and uncompensated. Although participation represented a small fraction of the current students, we gained important insights from the instrument.



\section{Collected innovations}

 While our findings are consistent with pedagogical principles that were understood pre-pandemic, several pedagogical changes have been normalized fruitfully by a year of necessary changes. We have grouped the innovations mentioned in the interviews into three categories and included quotations from student survey responses where relevant.

\subsection{Innovations in instructor-directed learning}

\subsubsection{Pre-pandemic}
Most classes before the pandemic used a traditional lecture format, meeting two to three times per week for 50-80 minutes each. Typically, student TAs were assigned at a ratio of 20-30 students per assistant, with an additional head TA allocated for larger courses over a few hundred students. TAs would be involved in homework preparation, discussion or lab sections, office hours providing direct student contact, and grading. Some courses were also partially supported by a professional course coordinator who oversaw management aspects of the course.  

For decades, select large-classroom lectures from this department have been recorded on video and made available online. With lecture material available on demand, residential student attendance in lecture often dropped to around 15\% after the third week of class. Office hours were often over-crowded, with long lines.

\subsubsection{Pandemic era}
The main innovations explored were:
\begin{itemize}
\item Disseminating content asynchronously through pre-recorded videos, available on demand 
\item Designing synchronous class time to complement pre-recorded material, including
\begin{itemize}
    \item Fireside chats with guest speakers, high-level overviews,and Q\&A
    \item Problem sessions, often dividing the class in smaller groups or smaller sections
    \item Real-time questions answered by course assistants over chat or online discussion platforms
\end{itemize}
\item Making synchronous lectures interactive with tools such as \textbf{\textit{ohyay}} \cite{ohyay}
\end{itemize}

\subsubsection{Instructor and student response}
\begin{figure*}

\includegraphics[width=0.9\textwidth,trim=0 0in 0 0in, clip]{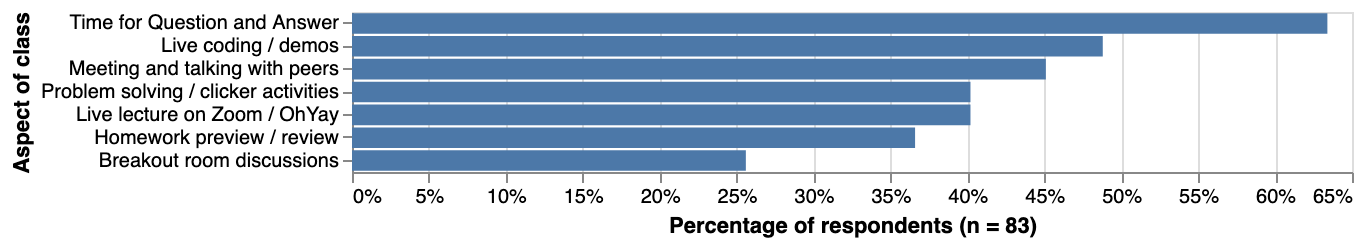}
\caption{Survey responses to "What aspects of synchronous class meetings did you appreciate? Check all that apply."}
\label{sync-sessions-fig}
\end{figure*}

Students gave synchronous sessions mixed reviews, as shown in
Figure~\ref{sync-sessions-fig}.
When synchronous sessions were recorded, we often saw the same low attendance 
as in our pre-pandemic in-person lectures.
A limitation of our survey is that it combines students from different classes, and therefore not all students surveyed were exposed to the same course components in the same degree.

A few instructors who were redesigning synchronous online lectures began to think of the experience as \textit{“more akin to going to a music performance.”} The instructors who used \textbf{\textit{ohyay}} found that their attendance was higher in the latter weeks of the course than it normally was when they were teaching in-person, suggesting that there is \textit{“still value to the synchronous lecture”} and in fact \textit{“there are some aspects to better educating people in this [online] space.” }\cite{fatahalian_2021}

Eighty-five percent of students responding to the survey 
said that it was “helpful” or “very helpful” to have  lecture material pre-recorded and 
available at any time. This suggests strong student demand for video recording in the future but does not necessarily suggest that courses should be entirely asynchronous.

In fact, students commented on the value of combining asynchronous and synchronous learning: \textit{"I really enjoyed having some asynchronous material and then having synchronous lectures that delved deeper through real-world examples once the concepts were out of the way."} Other students spoke about the ability to ask online/chat question during class: \textit{"TA and peer q and a during lectures was infinitely helpful."} Another echoed the importance of active learning: \textit{"The in-class problem solving activities in [course redacted] were great because we could anonymously message [name redacted] through zoom chat and then through what he said see what others are thinking/feeling."}

These remarks highlight the importance of using synchronous class meetings as a dynamic time for students to interact with their instructors to dive deeper into topics that they were already somewhat familiar with based on the pre-recorded lectures. The ability for students to ask more questions of their instructors (particularly anonymously via Zoom chat) in real time, potentially using the TAs to facilitate the Q\&A, was an important innovation that moves away from the previous model of largely one-way lectures. 

For some courses, TA-led small group sections provided a crucial opportunity for synchronous learning in the virtual setting. This builds on previous research that underscores the importance of near-peer mentors as approachable instructors \cite{10.1145/3017680.3017694}. Student remarks about these discussion sections included: \textit{"The only synchronous parts of class were section, which was helpful for reviewing content and/or getting elaborations on stuff I didn’t understand."} Additionally, another student said: \textit{"I think it is valuable to have sections meet synchronously. It was really valuable to interact with my peers and digest the material learned together....As people started to know
each other, the activities and interactions became more and more meaningful."}

\subsection{Innovations in student assessment}
\subsubsection{Pre-pandemic.}
Most classes before the pandemic used a combination of graded homework, quizzes, programming projects, midterm exams and cumulative high-stakes final exams. Grading has traditionally been competitive at this institution, with students feeling considerable pressure and anxiety even though grade curves often lead to a substantial number of high final grades. 


\subsubsection{Pandemic era.} University policy in spring 2020 required that all classes be taken pass/fail, and the 2020-2021 school year prohibited high-stakes final exams and eliminated the final exam week, although it should be recognized that some instructors replicated their traditional exam structures despite official policies. The primary innovations observed during 2020-21 were:
\begin{itemize}
    \item Multiple lower-stakes exams in lieu of one final exam
    \item Revise and resubmit, allowing students to revise submitted work to receive a higher grade (also known as mastery-based learning) 
    \item Pass/Fail grading policies
    \item Partner or group assignments and exams
    \item Concept checks, used regularly or weekly to track student academic progress and personal wellbeing
\end{itemize}

\subsubsection{Instructor and student response} Below, we consider the responses to each type of innovation: 

\textbf{Rethinking final exams.}
The remote learning environment prompted instructors to recognize the unequal access to stable learning environments and resources while students were under lockdown in their respective homes. For that reason, instructors who traditionally relied on final exams pivoted towards lower-stakes exams or a greater emphasis on assignments and projects. 

One instructor explained that he would not use any exams during the pandemic so as not to amplify disadvantages; he updated the grading policy to focus on assignment grades instead. As he said: \textit{"Right now... there are students who are not outfitted to be able to take exams ... Assuming that the people who are at a disadvantage and taking exams right now are probably not the people who would normally get top 5\% on an exam, that they're bringing a lot of their shortcomings, even to campus,...to actually amplify that disadvantage is... just not right."} He provided optional exams, which still gave students the opportunity to self-assess their knowledge.

When surveyed, 70\% of students considered it beneficial or very beneficial to have multiple smaller exams in lieu of one final exam, and 69\% of students surveyed found partner / group assignments beneficial or very beneficial. Students voiced appreciation for the changes above, which reduced stress: \textit{"It was helpful to not have exams in [course redacted] since I often struggle with exams but definitely understood the material based on the self-assessments and the assignments, so it just removed a lot of unnecessary stress."}
Some instructors who continued to give exams chose to accommodate more students by giving a 24-hour window to complete exams or by giving shorter exams. By giving shorter exams, some instructors also hoped to dissuade students from cheating, which remained a concern in the virtual learning environment. Two CS1 instructors reflected: \textit{"To mitigate [cheating] we only gave them 25 minutes so it's ...not enough time for you to cheat."}

\textbf{Revise and resubmit.}
For other instructors, the remote learning environment proved to be an opportunity to pilot mastery-based learning, which includes policies that allow students to revise and resubmit until they felt that they had achieved mastery of what they learned. As one instructor explained: \textit{"I've always wanted to experiment with some form of mastery learning, and this was my chance! We have 3 take-home exams, and after each take-home exam is graded, students whose work was unsatisfactory have an opportunity (with extensive feedback and staff support, though not the actual solutions) to revise and resubmit their work a week later. Students whose work is still unsatisfactory can revise \& resubmit again, with more staff support."}

Students responded favorably to these policies: 75\% of students who were in a course that offered the option to revise and resubmit work considered it beneficial or very beneficial to have the option to do so. The following comments from students spoke to the advantages of a mastery-based approach using a revise and resubmit policy: 
\textit{"[Instructor redacted]'s revise and resubmit policy in [course redacted] was fantastic. It meant I could engage with topics until I reached mastery, rather than just settling for 'good enough.'”} The sentiment was echoed from other students: \textit{"My learning experience feels so much more whole now that I’ve worked out the correct answer to every single midterm problem; I know this method of revision might be controversial because it doesn’t provide a memorization-based meritocracy grade distribution, but I feel like the CS program here has explained it’s goal to be to make all of us excellent programmers through hard work - which I think rewards revision, getting us students back on their feet after what might’ve been a demoralizing fall from grace.}

\textbf{Pass / Fail Grading Policies.}
The ability to take courses for a Pass/Fail grade instead of a letter grade changed students’ attitudes towards assessment. Teaching team members of the CS1 courses reflected on why Pass-Fail might be a helpful policy to level the playing field for introductory CS courses: \textit{"I think it's  not completely fair for intro CS classes to not be pass fail because there are people who come in with different backgrounds ... there are people who start [CS1] who had never coded before in their life and I have students who ... had coded since they were in middle school."}

\textbf{Weekly concept checks.} A number of courses created short quizzes each week for students to assess their understanding. 56\% of students surveyed who had weekly "concept checks" in their courses considered it beneficial or very beneficial to do them. When asked to elaborate on the use of “concept checks” to accompany the asynchronous lectures in a flipped classroom environment, one student stated: \textit{"Weekly concept checks in [course redacted] also helped me stay on track with the asynchronous lecture videos, and I thought that was a great idea because I never got behind on lecture which I often do."}

Student appreciation for each of the assessment methods is reflected in Figure 2. 
Because some of the innovations were not widely deployed, a number of respondents were not able to 
rate some of the methods, leading to lower scores than might have been seen otherwise. 

\begin{figure*}
\begin{tabular}{|l|r|r|r|r|r|}
\hline
&
Very beneficial &
Beneficial  &
Somewhat ben. &
Slightly ben. &
Not ben. \\
\hline
Option to revise and resubmit work  &
51.5\% &
23.53\% &
7.4\% &
11.8\% &
5.9\% \\
\hline
Weekly concept checks &
30.6\% &
25.88\% &
20.0\% &
11.8\% &
11.8\% \\
\hline
Multiple smaller exams in lieu of one final exam &
39.8\% &
30.11\% &
11.8\% &
10.8\% &
7.5\% \\
\hline
Partner / group assignments or exams &
50.6\% &
18.52\% &
12.3\% &
3.7\% &
14.8\%
\\
\hline
\end{tabular}
\caption{Student response to assessment options}
\label{student-assessment-table}
\end{figure*}

\subsection{Innovations in student-initiated learning}
\subsubsection{Pre-pandemic.}
Prior to the pandemic, office hours for larger lower-division classes were traditionally held in a designated location on campus multiple times throughout the week. The space, referred to as “the basement,” was often full of student activity, with spontaneous group formation, peer help and small group instruction with skilled TAs. Many students also had peers they could rely on to work with through class connections or other relationships they had built (e.g. student clubs). 

\subsubsection{Pandemic era.}
In the shift to remote instruction, these student-initiated learning options were arguably the hardest area to replace. 
Many of the ways that residential life supports student learning were taken for granted before the pandemic. 
Because the organizational structure of student peer collaboration was largely invisible to instructors, 
many instructors began the pandemic with little concrete understanding of how to replace residential peer learning. 
Because there was no single solution for this complex problem, iterative experiments with a range of platforms and techniques continued through the 2020-21 academic year, including:

\begin{itemize}
\item Discussion forums, allowing students to ask questions and other students or course staff to answer online
\item One-on-one office hours with course staff, as either drop-in, or 15 minute appointments
\item Working office hours, allowing students to join online rooms to work with other students, generally without course staff
\item "Homework Parties," in which students can gather remotely to meet and work together on \textbf{Nooks}\cite{nooks}, with course staff present at each session to provide help
\end{itemize}
Unlike in the instructor-initiated activities and the assessments, experiments in student-initiated learning struggled to replicate 
many of the in-person benefits of office hours and informal campus study options. 
However, a number of these formats were highly beneficial to the students that chose to take advantage of them. 
This suggests that hybrid structures supporting student collaboration might lead to productive innovations in the future. 

\subsubsection{Instructor and student response}
Despite the challenges of hosting online student-initiated activities, it is evident that students want to keep the option of online supports in the future: \textit{"Online OH [office hours] are so much more accessible and I actually feel like I got a lot more help this year than in previous years. [The basement] is chaotic and CAs are often running around like crazy and attention is always divided. [Course redacted] group office hours functioned really well online and my other classes ... had individual online OH that worked super, super well."} One graduate student compared the online option to the in-person experience: \textit{"Virtual OH with queues are, in my opinion, the best thing about remote courses and I hope continue when things go back to normal... the virtual queues on Nooks and Zoom made the process way more equitable ... And probably most crucially, if OH were crowded and there was a long queue, you could just stay on the queue and comfortably do other work from your own home or location as you waited to get helped."}



A number of students called out the benefits of using the \textbf{\textit{Nooks}} platform as a preferable alternative to \textbf{\textit{Zoom}} for meeting and collaborating with groups of peers during office hours. Finally, other students expressed appreciation for the active role that instructors took on in online discussion spaces: \textit{"[Instructor redacted] had a Slack workspace and he and the TA's were super active on it. I felt like I could receive and offer help to classmates, and I really felt like a part of a community."}

\section{Synthesis and trajectories}
\label{insightes-and-recommendations-sec}

In this section, we synthesize one or two main departmental priorities for each of the three categories and 
connect them to trends reported in the broader CS education literature.

\subsection{Instructor-directed learning insights}
The case study of our department suggests four key takeaways:
\begin{enumerate}
    \item Student appreciate when instructors make asynchronous lessons available prior to synchronous meetings because it allows students to get better acquainted with the material and maximizes the learning in synchronous sessions. 
    \item Synchronous class time is most beneficial to students when designed around “active learning” activities such as live problem practice, question and answers with the teaching team, fireside chats with guest speakers, and extending course material to apply it to real-world scenarios.
    \item Using technology to enable more opportunities for students to interact with the teaching team and with peers makes learning more visible and creates more opportunities for important discussions.
    \item Instructors who collect frequent student feedback using surveys and check-ins have a better pulse on how their students are doing in the class and in their personal lives, and this empathy translates to a better learning experience for students.
\end{enumerate}
Two of these changes speak to shifting trends over the past decade of CS education that feel useful to elaborate upon. 

\textbf{Making asynchronous videos available.}
One clear trend is that students strongly value the scheduling flexibility of asynchronous video presentations. The ability to watch short lecture videos prior to synchronous course meetings increases the value of the time spent as a group. Also, other institutions have reported that the ability to rewatch and refer back to recordings of synchronous meetings is useful for students \cite{culbert_2021}. These practices are a nod to previous experiments with flipped classroom models \cite{10.1145/2591708.2591752}. While the research suggests that there is often low compliance to watch asynchronous videos \cite{10.1145/2899415.2899468}, the recordings might still be effective if they are paired with concept checks and/or if the videos' relevance to synchronous meetings is properly communicated.  

\textit{Recommendation:} Make (short) pre-recorded videos available so that students can cover the material at their own pace and can be better prepared for synchronous meetings. This can also reduce the number of synchronous class meetings per week.

\textbf{New models of active learning in CS.}
Active learning teaching practices in Physics \cite{crouch2001peer}, CS \cite{10.1145/2445196.2445248}, and other STEM disciplines have been around for quite a long time. While many instructors were hesitant to change their practices and move away from traditional lectures prior to the pandemic, the shift to remote instruction caused many to rethink how to best use synchronous meetings. The experiments at our institution did not reveal one “best way” to engage students during synchronous class meetings. However, the four quarters of instruction did reveal that students preferred classes in which they were (1) actively solving practice problems, (2) asking questions of the instructor and the TAs (via chat or audio), and (3) able to learn about real-world applications of relevant course topics, including from guest speakers who are professionals in the field. 

These findings suggest that “active learning” might need to be more explicitly defined in CS to encompass a broad range of activities in which students are engaged with the instructor, the TAs, their classmates and/or guest speakers during synchronous meetings. Making the course responsive to student needs and questions, while creating opportunities to take advantage of the precious human resources and opportunities for social interaction appears to greatly benefit learning \cite{10.1145/2445196.2445248,sarma2020grasp}.

\textit{Recommendation:} Instructors should structure synchronous class time to engage students in activities, dialogue and practice problems that make learning more social and provide opportunities for human interaction. Instructors should survey their students and collect frequent feedback so that they can continue to experiment and iterate with new formats for “active learning” methodologies that are best suited to CS instruction in their courses.  

\subsection{Assessment insights}

The choice by many institutions to require Pass/Fail grading in spring 2020 in response to the pandemic forced instructors and students to approach grades and assessment differently than they had in the past. Some students found it liberating, while others were stressed about not knowing how to get an A in the new class formats. The experiments in formative and summative assessments highlighted a few key points of conversation within our department: 
\begin{enumerate}
     \item Formative and summative assessments can take many forms. High-stakes exams are not the only way to assess student learning, and more frequent exams and concept checks might help students to assess how they are doing throughout a course. Projects, group assignments and homeworks can all help to illuminate student understanding.
    \item Revise and resubmit policies that support mastery of course material reduce student stress and provide powerful opportunities for student learning and success. A forced grading curve may not be necessary, particularly in introductory courses. 
    \item Some students appreciate grades and need graded transcripts to apply for jobs and future degrees, while others are more focused on learning. 
    \item Online exams are susceptible to cheating and misuse.
\end{enumerate}

\textbf{Formative and summative assessments take many forms.}
Important innovations from the past experiments at our institution included the use of frequent “concept checks” for students to self-assess their understanding of the material as they moved through the course, rather than having to wait for a higher-stakes midterm or final to find out how well they understood the content. Similarly, more frequent but lower-stakes summative exams seemed to reduce the stress on students and allow for faster learning cycles, although more research is warranted. At other institutions, alternative assessments proliferated as well, such as open-book exams \cite{10.1145/3430665.3456373}, group assessments \cite{10.1145/3408877.3432412} and oral exams \cite{10.1145/3408877.3432511} with generally positive results. 

\textit{Recommendation:} Use frequent assessment to reduce student stress and scaffold learning effectively.

\textbf{Mastery-Based Learning policies such as “revise and resubmit” support student agency \& learning.}
Students who had the opportunity to “revise and resubmit” homework and assessments found that they were better able to learn the content because they felt supported in continuing to learn the material until they proved they had mastered it. Similar approaches, often termed mastery-based or competency-based learning in education research \cite{voorhees2001competency}, has been shown to improve student outcomes by shifting the focus away from exams to learning. The incredible response by students and instructors alike to these policies suggest that both groups see tremendous benefits in this approach, with opportunities to support learners without putting too much of a burden on instructors \cite{leetalk}. 

\textit{Recommendation:} Consider adopting policies that support mastery learning methods.

\subsection{Student-initiated learning insights}
Shifting office hours and other essential yet often overlooked peer learning structures online was a major challenge at our institution. The ability to recreate the types of support systems that informally and spontaneously arose in in-person settings was not readily available online. However, new systems did emerge to reach students, especially those struggling in courses. Key insights included: 

\begin{enumerate}
    \item Both in-person and online office hours benefit students, and it is important to have both options going forward as certain formats benefits certain types of students. 
    \item Creating opportunities for students to collaboratively work together and help each other should be more deliberately designed for going forward. Leaving it to chance advantages certain students (e.g. students with more friends in the major, students who are more comfortable asking questions in class or in office hours, etc.).

\end{enumerate}

Looking to the future, it is evident that students at our institution are excited to be back on campus in proximity to their peers and with access to in-person academic spaces. However, the role of online support systems should not be overlooked as a key resource for many students. 

\textbf{Recognize that students have different prior experiences;  meet students where they are.}
The strong desire to both return to in-person office hours and to keep online office hours, as well as other new structures such as “homework parties” suggests that different students learn in different way and need different supports. For some who don’t feel comfortable in hectic environments or in-person interactions, online office hours will continue to be their preferred method for seeking help. For students that thrive in group environments, an in-person setting might be preferred. For students without strong social connections in their classes, facilitated “homework parties” can provide invaluable access to peers. This takeaway echoes Universal Design for Learning principles \cite{rose2002teaching} and other peer-learning models from the pandemic \cite{pprspeertalk}.

\textit{Recommendation:} 
Meet students where they are by creating multiple formats for students to get support. Consider different learning needs in both online and in-person formats.

\section{Conclusions and Future Work}
A descriptive case study of four academic quarters from spring 2020 to spring 2021 revealed energetic innovation and iterative experimentation, and student comments strongly suggest that student expectations and demands have been changed dramatically by this period. Remote learning at this institution has at least partially continued to the time of writing. 
Given instructor investment in the work they have done over this period, and the receptiveness of students to many of these innovations, the most promising steps forward will draw productively on the 2020-21 experience. 
Specific suggestions are given in section \ref{insightes-and-recommendations-sec}, and future research is necessary to support the following trends: 

\textit{Trends in instructor-directed learning.} Based on current and prior experience, it is virtually certain that  pre-recorded video and class session recordings will continue
to be valuable. However, with the low student attendance that occurs when synchronous online sessions are recorded and accessible on demand, 
there is a need to thoughtfully design synchronous course activities to foster productive student engagement.

\textit{Trends in student assessment.} During the pandemic, some instructors provided more flexible deadlines to accommodate students in challenging circumstances, replaced their single high-stakes final exam with alternatives, and explored mastery-based learning. However, we predict that a nebulous transition period will be necessary for all instructors and students to co-create and fully adopt a new academic culture surrounding assessment.

\textit{Trends in student-initiated learning.} Many of the ways that residential life supports student learning were taken for granted before the pandemic but became shockingly apparent as soon as we were confronted with their loss. Because the organizational structure of student peer collaboration was largely invisible to instructors, instructors began the pandemic with little concrete understanding of how to replace residential peer learning.  As there was no single solution for this complex problem, iterative experiments with a range of platforms and techniques continue.

A clear challenge for this department is to solidify and effectively expand the use of pedagogy that was ushered in by the pandemic. We hope that the simplicity of the recommendations highlighted in this paper will be useful in focusing attention on a few selected incremental changes that are consistent with current instructor and student views. Beyond those incremental steps, we hope further that by encouraging instructors to monitor how well they support students during the pandemic, we may have opened the door to longer-term interest in adopting research-driving pedagogical practices that are consistent with the local culture and values of the institution.

\section{Acknowledgements}
We extend our deepest gratitude to the department's instructors who responded valiantly to the challenges of the past years with tremendous dedication to their students. We are indebted to the instructors and students who participated in our case study. 

\bibliographystyle{ACM-Reference-Format}
\balance
\bibliography{sample-base}

\end{document}